\documentclass[12pt]{iopart}
\usepackage{iopams}  
\usepackage{graphics}
\usepackage{epsfig}
\usepackage{cite}
\eqnobysec

\def\lambdabar {\mathchar'26\mkern-10mu\lambda}

\begin{document}

\title[The Wigner d.f. of the relativistic finite-difference oscillator in an external field]{On the Wigner function of the relativistic finite-difference oscillator in an external field}

\author{S.M. Nagiyev, G.H. Guliyeva and E.I. Jafarov\footnote[3]{E-mail: ejafarov@physics.ab.az}}

\address{Institute of Physics, Azerbaijan National Academy of Sciences, Javid av. 33, AZ1143, Baku, Azerbaijan}

\date{\today}% It is always \today, today,

\begin{abstract}
The phase-space representation for a relativistic linear oscillator in a homogeneous external field expressed through the finite-difference equation is constructed. Explicit expressions of the relativistic oscillator Wigner quasi-distribution function for the stationary states as well as of states of thermodynamical equilibrium are obtained and their correct limits are shown.
\end{abstract}

\pacs{03.65.Pm, 02.30.Gp, 03.65.Vf}

\submitto {\JPA}

%\keywords{phase space; relativistic finite-difference oscillator; Wigner function}
%Use showkeys class option if keyword
%\maketitle

%%%%%%%%%%%%%%%%%%%%%%%%%%%%%%
%%%   SECTION              %%%
%%%%%%%%%%%%%%%%%%%%%%%%%%%%%%
\section{Introduction}

The problem of a one-dimensional harmonic oscillator is one of the important paradigms of the theoretical physics. Its solution in the classical approach is unique and simple, leading to the enormous applications in a wide range of the modern physics~\cite{moshinsky}. This problem in the non-relativistic quantum approach has at least the same importance, and probably the main reason is that it has very elegant solutions to the Schr\"odinger equation in terms of the well-known Hermite polynomials and the algebra of this problem can be easily factorized as Heisenberg algebra with its generators being the starting point of the quantum field theory. Here, one needs to note that there is another important solution to the Schr\"odinger equation in terms of the Hermite polynomials and it is the non-relativistic harmonic oscillator in an external homogeneous field~\cite{landau}. The importance of this non-relativistic problem can also be observed through its numerous applications. For example, the theory of the Brownian motion of a quantum oscillator is developed by using this model~\cite{agarwal}, and a model of a harmonic oscillator in an external gravitational field is considered and the developed formalism is applied to the study of thermal properties of noninteracting Bose and Fermi gases in harmonic traps~\cite{wallis,kulikov,saif}. It is necessary to note the generalization of this problem to the two-dimensional case, where an elastically bound particle in a space with a combined linear topological defect is studied in detail~\cite{azevedo1,azevedo2}.

However, the solution to the problem under consideration  in the configuration space is not enough and an important step here is to study the quantum problem with correspondence to its classical analogue. The phase-space approach allows us to answer this question and opens a lot of hidden features of the quantum system expressing it in the language of the classical approach. In order to see them, we just need to compute the quasi-probability function of the joint distribution of the momentum and position, where the Wigner distribution function is most known. Analysing the Wigner distribution function for the non-relativistic harmonic oscillator in an external field, we can observe that the applied linear field just shifts its stationary states to the negative position direction and does not have any influence to the values of the momentum. In this case, an important question about any role of the relativistic effect arises. The purpose of this paper is the phase-space study of the one-dimensional relativistic oscillator model in an external homogeneous field expressed through finite-difference Schr\"odinger-like equation~\cite{atakishiyev-nonlin,atakishiyev-adv,atakishiyev-rep}.

Our paper has following structure. In section 2, we give brief information about the Wigner distribution function. The Section 3 is devoted to the non-relativistic linear oscillator in an external field, where we present its solutions both in configuration and phase spaces as well as the expression of the Wigner function of the thermodynamic equilibrium. Section 4 is devoted to the model of the finite-difference relativistic oscillator in an external field and in section 5, we present the explicit expressions of the Wigner distribution function of the stationary and thermodynamic equilibrium states for the relativistic linear oscillator in an external field.

%%%%%%%%%%%%%%%%%%%%%%%%%%%%%%
%%%   SECTION              %%%
%%%%%%%%%%%%%%%%%%%%%%%%%%%%%%
\section{The Wigner quasi-probability distribution function}

The Wigner function $W\left( {p,x;t} \right)$~\cite{wigner} being as analogue of the classical distribution function in the phase space $\rho \left( {p,x} \right)$ has a wide range of applications in the non-relativistic quantum mechanics~\cite{lee,hillery,tatarskii}. The limit relation $\mathop {\lim }\limits_{\hbar  \to 0} W(p,x;t) = \rho \left( {p,x} \right)$ between the Wigner function and the classical distribution function exists and it allows us to calculate quantum corrections to the known classical results employing an analytical form of the Wigner function. It is necessary to note that the first application of this function was the calculation of the quantum corrections to the classical distribution function of the equilibrium states of particle system in an arbitrary potential field. The Wigner d.f. is a function of momentum $p$ and position $x$ as well as a fucntion of the time $t$ in a general case. One can obtain it from the wavefunctions of the quantum system under consideration both in position representation $\psi \left( x \right)$ and momentum representation $\phi \left( p \right)$ by using well-known transitions:

\begin{eqnarray}
\label{2.1}
 W\left( {p,x;t} \right) = \frac{1}{{2\pi \hbar }}\int\limits_{ - \infty }^\infty  {\psi ^* \left( {x + \frac{1}{2}x',t} \right)e^{ ipx'\hbar } } \psi \left( {x - \frac{1}{2}x',t} \right)dx', \\ 
\label{2.2}
 W\left( {p,x;t} \right) = \frac{1}{{2\pi \hbar }}\int\limits_{ - \infty }^\infty  {\phi ^* \left( {p + \frac{1}{2}p',t} \right)} e^{ - ixp'/\hbar } \phi \left( {p - \frac{1}{2}p',t} \right)dp'.
\end{eqnarray}

The Wigner function (\ref{2.1})-(\ref{2.2}) satisfies the following equations:

\begin{eqnarray}
\label{2.3}
 \int {W(p,x;t)dp = \left| {\psi (x,t)} \right|} ^2  = W(x,t), \\ 
\label{2.4}
 \int {W(p,x;t)dx = \left| {\phi (p,t)} \right|} ^2  = W(p,t).
\end{eqnarray}

Here, $W(x,t)$ is the probability of the particle observation at point $x$ at time $t$. Correspondingly, $W(p,t)$ is the probability of the particle observation in the momentum space with the momentum value $p$ at time $t$.

Despite the fact that the Wigner function $W\left( {p,x;t} \right)$ satisfies equations (\ref{2.3}) and (\ref{2.4}), one cannot consider it as the probability of the observation of the particle with the momentum value $p$ at point $x$ due to the fact that the function $W\left( {p,x;t} \right)$ at some values of the $p$ and $x$ can become negative.

With the help of the Wigner function one can find the average value of any physical parameter $f(p,x)$ through the following formula:

\begin{equation}
\label{2.5}
\bar f = \int {f(p,x)W(p,x,t)dpdx} .
\end{equation}

An explicit expression of the Wigner function has been already obtained for a number of the non-relativistic quantum mechanics problems~\cite{lee,hillery,tatarskii,davies,akhundova}, whereas the phase space of the relativistic model of the linear oscillator described by the finite-difference equation~\cite{atakishiyev1} was considered in~\cite{atakishiyev-wigner}.

The purpose of this paper is to obtain the explicit expression of the Wigner function for the relativistic model of the linear oscillator in an external homogeneous field.

%%%%%%%%%%%%%%%%%%%%%%%%%%%%%%
%%%   SECTION              %%%
%%%%%%%%%%%%%%%%%%%%%%%%%%%%%%
\section{The non-relativistic linear oscillator in an external field}

Let us first to consider in short the non-relativistic case. In the non-relativistic quantum mechanics, the Hamiltonian of the linear oscillator in a homogeneous external field

\begin{equation}
\label{3.1}
H_N^g  =  - \frac{{\hbar ^2 }}{{2m}}\frac{{d^2 }}{{dx^2 }} + \frac{{m\omega ^2 x^2 }}{2} + gx
\end{equation}
has the following eigenfunctions~\cite{landau}:

\begin{equation}
\label{3.2}
\psi _{Nn}^g (x) = c_{Nn}  \cdot H_n \left( {\left( {x + x_0 } \right)\sqrt {\frac{{m\omega }}{\hbar }} } \right) \cdot e^{ - \frac{{m\omega }}{{2\hbar }}(x + x_0 )^2 } .
\end{equation}

These wavefunctions correspond to the energy spectrum

\begin{equation}
\label{3.3}
E_{Nn}^g  = \hbar \omega \left( {n + \frac{1}{2}} \right) - \frac{{m\omega ^2 }}{2}x_0^2 
\end{equation}
with $x_0  = g/m\omega ^2 $. Function (\ref{3.2}) satisfies the following orthonormalization relation:

\begin{equation}
\label{3.4}
\int\limits_{ - \infty }^\infty  {\psi _{Nn}^{g*} (x)\psi _{Nm}^g (x)dx}  = \delta _{nm} .
\end{equation}

From (\ref{3.4}) it follows that

\[
c_{Nn}  = \frac{{c_{N0} }}{{\sqrt {2^n n!} }},\quad c_{N0}  = \left( {\frac{{m\omega }}{{\pi \hbar }}} \right)^{1/4} .
\]

One can obtain the wavefunctions $\psi _{Nn}^g (x)$ (\ref{3.2}) by the simple transition from the wave functions $\psi _{Nn}^0 (x)$ of the non-relativistic oscillator at $g=0$, i.e.:

\[
\psi _{Nn}^g (x) = e^{ - i\hbar x_0 \frac{\partial }{{\partial x}}} \psi _{Nn}^o (x) = \psi _{Nn}^0 (x + x_0 ).
\]

In the momentum representation this transition is as follows:

\[
\phi _{Nn}^g (p) = e^{\frac{{ix_0 p}}{\hbar }} \phi _{Nn}^0 (p).
\]

After substitution of (\ref{3.2}) into (\ref{2.1}) one can perform the integration and we will find the Wigner function of the non-relativistic linear oscillator stationary states in an external field (\ref{3.1}):

\begin{equation}
\label{3.5}
W_{Nn}^g (p,x) = \frac{{( - 1)^n }}{{\pi \hbar }}e^{ - \left( {\eta ^2  + (\xi  + \xi _0 )^2 } \right)} L_n \left( {2\eta ^2  + 2(\xi  + \xi _0 )^2 } \right),
\end{equation}
where $\eta  = \frac{p}{{\sqrt {m\omega \hbar } }}$ and $\xi  = x\sqrt {\frac{{m\omega }}{\hbar }} $ are the dimensionless variables, $\xi _0  = x_0 \sqrt {\frac{{m\omega }}{\hbar }}  = \frac{g}{\omega }\sqrt {\frac{{m\omega }}{\hbar }} $ is the dimensionless parameter and $L_n \left( x \right)$ is the Laguerre polynomial.

By using equation (\ref{2.2}), one can obtain the following operator form of (\ref{3.5}):

\begin{equation}
\label{3.5a}
W_{Nn}^g \left( {p,x} \right) = \frac{1}{{2^n n!}}\frac{1}{{\pi \hbar }}H_n \left( {\eta  + \frac{i}{2}\partial _\xi  } \right) \cdot H_n \left( {\eta  - \frac{i}{2}\partial _\xi  } \right)e^{ - \left[ {\eta ^2  + \left( {\xi  + \xi _0 } \right)^2 } \right]} .
\end{equation}

It is necessary to note that the Wigner function (\ref{3.5}) is normalized by the condition

\[
\int\limits_{ - \infty }^\infty  {W_{Nn}^g (p,x)dpdx = 1} ,
\]
from where we obtain that

\begin{equation}
\label{3.6}
\sum\limits_{k = 0}^n {\frac{{\Gamma \left( {k + 1/2} \right)\Gamma \left( {n - k + 1/2} \right)}}{{k!(n - k)!}}}  = \pi .
\end{equation}

Another summation formula that we will obtain through the substitution of (\ref{3.2}) to (\ref{2.3}) is:

\begin{equation}
\label{3.7}
\sum\limits_{k = 0}^n {2^k C_n^k  \cdot \left( {2k - 1} \right)!! \cdot H_{2n - 2k} \left( {x\sqrt 2 } \right)}  = 2^n H_n ^2 \left( x \right),
\end{equation}
where by definition $0!! = \left( { - 1} \right)!! = 1$~\cite{prudnikov-I}. If we take into account the equality $\left( {2k - 1} \right)!! = \frac{{2^k }}{{\sqrt \pi  }}\Gamma \left( {k + 1/2} \right)$, then this formula can be written in an equivalent form as

\begin{equation}
\label{3.7b}
\frac{1}{{\sqrt \pi  }}\sum\limits_{k = 0}^n {2^{2k} C_n^k \Gamma \left( {k + 1/2} \right)H_{2n - 2k} \left( {x\sqrt 2 } \right)}  = 2^n H_n^2 \left( x \right).
\end{equation}

The Wigner function of the quantum system in the state of the thermodynamic equilibrium at temperature $T$ is determined by the formula

\begin{equation}
\label{3.8}
W_N^g (p,x) = \sum\limits_{n = 0}^\infty  {w_{Nn}^g W_{Nn}^g } (p,x),
\end{equation}
where

\begin{equation}
\label{3.9}
w_{Nn}^g  = \frac{{e^{ - \beta E_{Nn}^g } }}{{Z_N^g (\beta )}} = 2\sinh (\beta \hbar \omega /2)e^{ - \beta E_{Nn}^0 }  = w_{Nn}^0,\quad \beta  = 1/kT .
\end{equation}

It is possible to perform the summation (\ref{3.8}) for the case of the non-relativistic linear oscillator in an external field, where we will obtain an explicit expression of the equilibrium Wigner function as follows:

\begin{equation}
\label{3.10}
W_N^g (p,x) = \frac{{\tanh \left( {\beta \hbar \omega /2} \right)}}{{\pi \hbar }}\exp \left[ { - \left( {\eta ^2  + (\xi  + \xi _0 )^2 } \right)\tanh (\beta \hbar \omega /2)} \right].
\end{equation}

%%%%%%%%%%%%%%%%%%%
\begin{figure}
\begin{center}
\begin{tabular}{cc}
\includegraphics[width=0.35\textwidth,angle=270]{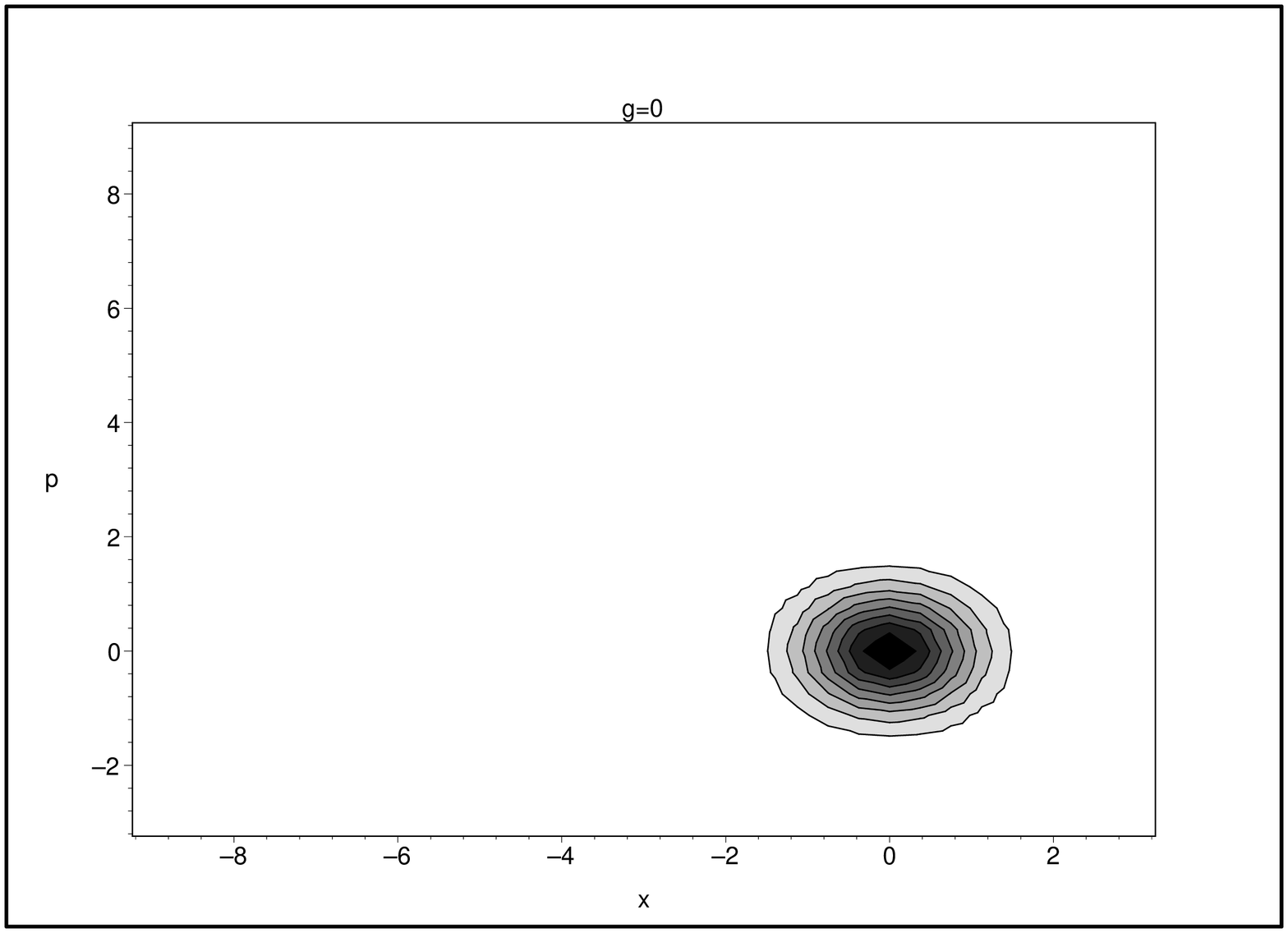}&
\includegraphics[width=0.35\textwidth,angle=270]{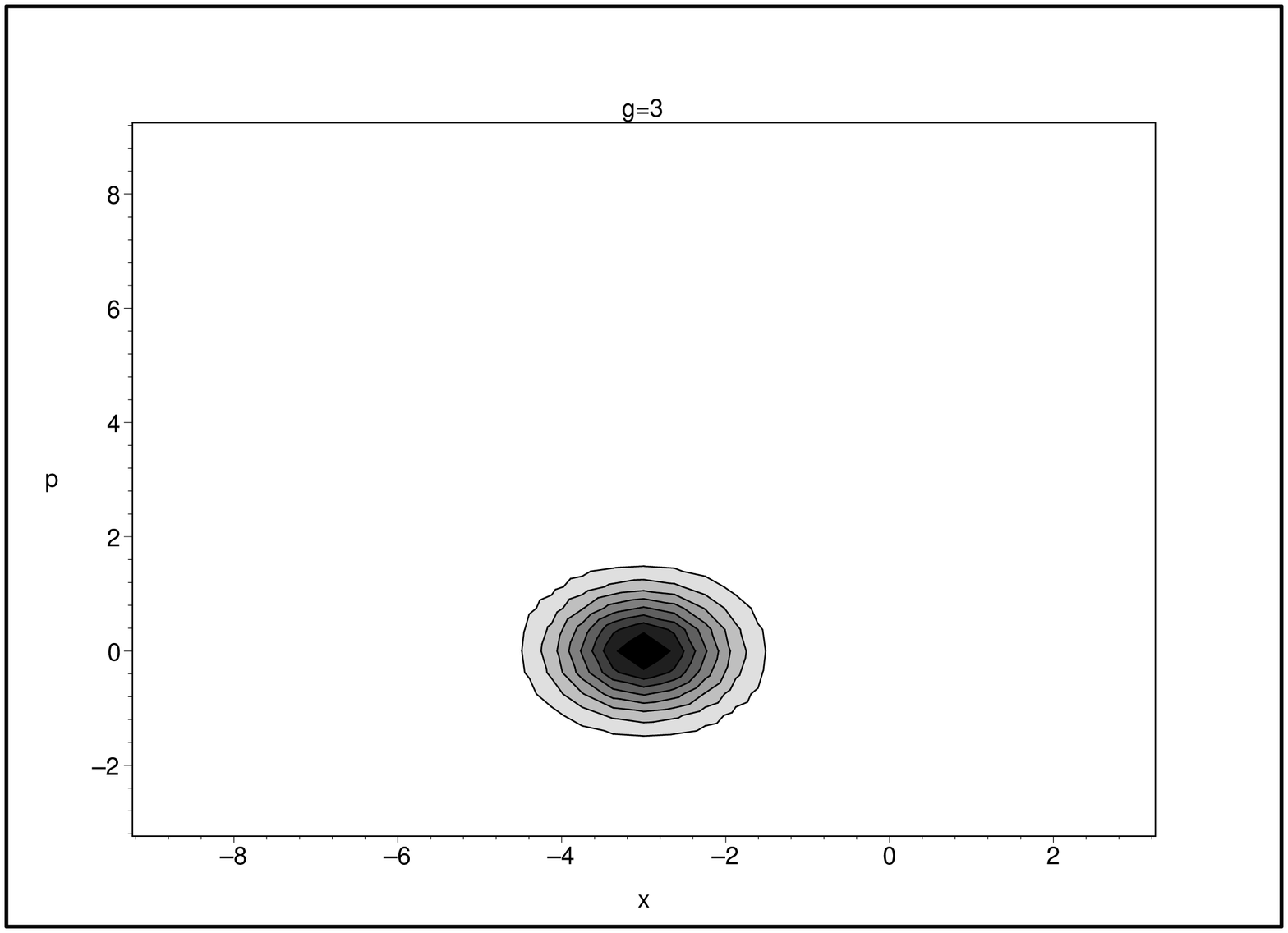}
\end{tabular}
\end{center}
\caption{\label{fig1} The behaviour of the ground-state Wigner function of the non-relativistic linear oscillator in an external homogeneous field for values of the external field $g=0$ and $g=3$ ($m=\omega=\hbar=1$)}
\end{figure}
%%%%%%%%%%%

The behavour of the ground state Wigner function $W_{N0}^g (p,x)$ determined by equation~(\ref{3.5}) is presented in \Fref{fig1}. On can see that the applied external field in the non-relativistic case does not influence the joint distribution form just shifting it along the negative values of the position.

%%%%%%%%%%%%%%%%%%%%%%%%%%%%%%
%%%   SECTION              %%%
%%%%%%%%%%%%%%%%%%%%%%%%%%%%%%
\section{The relativistic linear oscillator in an external field}

The relativistic model of the linear oscillator in an external homogeneous field is described by the finite-difference Hamiltonian~\cite{atakishiyev-nonlin,atakishiyev-adv,atakishiyev-rep}

\begin{equation}
\label{4.1}
H^g (x) = mc^2 \cosh i\lambdabar \partial _x  + \frac{{m\omega ^2 }}{2}x(x + i\lambdabar )e^{i\lambdabar \partial _x }  + gx,
\end{equation}
where $\lambdabar  = \hbar /mc$ is the Compton wavelength of the particle with mass $m$ and parameter $c$ is the speed of light.

Hence, one needs to mention that the dynamical symmetry group and Barut-Girardello coherent states were constructed for model (\ref{4.1}) and the bilinear generating function for the Meixner-Pollaczek polynomials is obtained in~\cite{atakishiyev-nonlin} and the generalized coherent states for the model of the relativistic linear oscillator were constructed in~\cite{atakishiyev-rep}.

The eigenfunctions corresponding to the Hamiltonian $H^g (x)$ (\ref{4.1}) have the following form:

\begin{equation}
\label{4.2}
\psi _n^g (x) = c_n^g \left( {\frac{{\hbar \omega }}{{mc^2 }}} \right)^{ix/\lambdabar } \Gamma \left( {\nu  + ix/\lambdabar } \right)P_n^\nu  \left( {\frac{x}{\lambdabar };\varphi } \right)e^{\left( {\varphi  - \pi/2 } \right)x/\lambdabar } ,
\end{equation}
where the normalization constant $c_n^g$

\begin{equation}
\label{4.3}
\fl \qquad c_n^g  = e^{in\left( {\pi /2 - \varphi } \right)} (1 - e^{ - 2i\varphi } )^\nu  \sqrt {\frac{{n!}}{{2\pi \lambdabar \Gamma (n + 2\nu )}}} ,\quad \nu  = \frac{1}{2} + \sqrt {\frac{1}{4} + \left( {\frac{{mc^2 }}{{\hbar \omega }}} \right)^2 } ,
\end{equation}
and $P_n^\nu  (x;\varphi )$ is the Meixner-Pollaczek polynomial.

The wavefunctions (\ref{4.2}) are orthonormalized by the following condition:

\[
\int\limits_{ - \infty }^\infty  {\psi _n^{g*} (x)\psi _m^g (x)dx = \delta _{nm} } .
\]

The energy spectrum of the relativistic oscillator in an external field corresponding to the Hamiltonian $H^g (x)$ (\ref{4.1}) and the wavefunctions (\ref{4.2}) has the following form:

\begin{equation}
\label{4.4}
E_n^g  = \hbar \omega \delta \left( {n + \nu } \right),\quad n = 0,1,2,3, \ldots ,
\end{equation}
where the spectrum (\ref{4.4}) becomes discrete under the condition $\left| g \right| < mc\omega $ and the angle $0 < \varphi  < \pi $ is determined through the relations $\cos \varphi  = g/mc\omega $ and $\delta  = \sin \varphi $.

One can obtain the wavefunctions $\phi _n^g \left( p \right)$ of the quantum system in the momentum representation from the wavefunctions $\psi _n^g (x)$  (\ref{4.2}) through the relativistic Fourier transform:

\begin{equation}
\label{4.5}
\phi _n^g \left( p \right) = \frac{1}{{\sqrt {2\pi \hbar } }}\int\limits_{ - \infty }^\infty  {\xi ^* (p,x)\psi _n^g (x)dx} ,
\end{equation}
where the function

\begin{equation}
\label{4.6}
\xi \left( {p,x} \right) = \left( {\frac{{p_0  + p}}{{mc}}} \right)^{ix/\lambdabar }  = e^{ix\chi /\lambdabar } 
\end{equation}
is the relativistic plane wave~\cite{atakishiyev1,shapiro} and $\chi  = \ln ((p_0  + p)/mc)$ is the rapidity. As a result, we find that the wavefunctions of the momentum representation are expressed through the generalized Laguerre polynomial:

\begin{eqnarray}
\label{4.7}
 \phi _n^g \left( p \right) = c_n ^\prime  \zeta ^\nu  e^{\gamma \zeta } L_n^{2\nu  - 1} (2\delta \zeta ), \\ 
 c_n ^\prime   = i^n \left( {2\delta } \right)^\nu  \sqrt {\frac{{n!}}{{mc\Gamma (n + 2\nu )}}} ,\quad \zeta  = \frac{{c(p_0  + p)}}{{\hbar \omega }} = \frac{{mc^2 }}{{\hbar \omega }}e^\chi  ,\quad \gamma = ie^{i\varphi } . \nonumber
\end{eqnarray}

%%%%%%%%%%%%%%%%%%%%%%%%%%%%%%
%%%   SECTION              %%%
%%%%%%%%%%%%%%%%%%%%%%%%%%%%%%
\section{Wigner representation of the relativistic oscillator in an external filed}

Let us now construct the Wigner representation of the relativistic oscillator in an external homogeneouf field. First, we will consider Wigner function of the oscillator stationary states and we will use for our calculations the following definition of the Wigner function for the relativistic case~\cite{atakishiyev-wigner,alonso}:

\begin{equation}
\label{5.1}
W_n^g (p,x) = \frac{1}{{2\pi \lambdabar }}\int\limits_{ - \infty }^\infty  {\phi _n^{g*} \left( {\chi  + \frac{1}{2}\chi '} \right)\phi _n^g \left( {\chi  - \frac{1}{2}\chi '} \right)} e^{ - ix\chi '/\lambdabar } d\chi ',
\end{equation}
where $\phi _n^g (p) \equiv \phi _n^g (\chi )$.

Substituting (\ref{4.7}) into (\ref{5.1}) and taking into account the relation

\[
e^{ \pm \frac{{i\lambdabar }}{2}\partial _x } e^{ - \frac{{2ix\chi '}}{\lambdabar }}  = e^{ \pm \chi '} e^{ - \frac{{2ix\chi '}}{\lambdabar }}  
\]
we will obtain the following finite-difference expression for the Wigner function of the stationary oscillator states in an external field:

\begin{equation}
\label{5.2}
W_n^g (p,x) = \frac{{n!}}{{(2\nu )_n }}L_n^{2\nu  - 1} (2\delta \zeta e^{\frac{{i\lambdabar }}{2}\partial _x } )L_n^{2\nu  - 1} \left( {2\delta \zeta e^{\frac{{ - i\lambdabar }}{2}\partial _x } } \right)W_0^g  \left( {p,x} \right).
\end{equation}

Now, we can obtain the explicit expression of the Wigner function of the relativistic oscillator ground state in an external field as follows:

\begin{equation}
\label{5.3}
W_0^g \left( {p,x} \right) = \frac{{(2\delta \zeta )^{2\nu } }}{{\pi \hbar \Gamma (2\nu )}}\int\limits_{ - \infty }^\infty  {e^{\gamma \zeta z + \gamma^* \zeta /z} z^{\frac{{2ix}}{\lambdabar } - 1} dz} .
\end{equation}

With the help of the following integral relation~\cite{prudnikov-I}

\begin{equation}
\label{5.3a}
\int\limits_0^\infty  {x^{\alpha  - 1} e^{ - px - q/x} dx = 2\left( {\frac{q}{p}} \right)^{\alpha /2} K_\alpha  \left( {2\sqrt {pq} } \right)} ,\quad {\mathop{\rm Re}\nolimits} p > 0, {\mathop{\rm Re}\nolimits} q > 0,
\end{equation}
one can express (\ref{5.3}) through the Macdonald functions:

\begin{equation}
\label{5.4}
W_0^g (p,x) = \frac{{2\left( {2\delta \zeta } \right)^{2\nu } }}{{\pi \hbar \Gamma \left( {2\nu } \right)}}e^{\left( {2\varphi  - \pi } \right)x/\lambdabar } K_{2ix/\lambdabar } \left( {2\zeta } \right).
\end{equation}

In a similar way, we obtain the following expression for the Wigner function of the excited states:

\begin{eqnarray}
\label{5.5}
W_n^g (p,x) = \frac{{2(2\delta \zeta )^{2\nu } n!}}{{\pi \hbar \Gamma \left( {n + 2\nu } \right)}}e^{\left( {2\varphi  - \pi } \right)x/\lambda }   \\ 
 \times \sum\limits_{k,j = 0}^n {\left( \begin{array}{l}
 n + 2\nu  - 1 \\ 
 n - k \\ 
 \end{array} \right)\left( \begin{array}{l}
 n + 2\nu  - 1 \\ 
 n - j \\ 
 \end{array} \right)\frac{{\left( { - 2\delta \zeta } \right)^{k + j} }}{{k!j!}}e^{i\left( {\varphi  - \frac{\pi }{2}} \right)\left( {k - j} \right)} K_{2ix/\lambdabar  + j - k} (2\zeta )},  \nonumber
\end{eqnarray}
where $\left( a \right)_n  = a\left( {a + 1} \right) \cdots \left( {a + n - 1} \right) = \Gamma \left( {a + n} \right)/\Gamma \left( a \right)$ is the Pochhammer symbol, $\left( \begin{array}{l}
 a \\ 
 n \\ 
 \end{array} \right) = \frac{{\Gamma \left( {a + 1} \right)}}{{n!\Gamma \left( {a + 1 - n} \right)}}$ are the binomial coefficients and here we also used the explicit form of the generalized Laguerre polynomials:

\[
L_n^a (x) = \sum\limits_{k = 0}^n {\frac{{\left( { - 1} \right)^k }}{{k!}}\left( \begin{array}{l}
 n + a \\ 
 n - k \\ 
 \end{array} \right)x^k } .
\]

The limit relation, reducing the (\ref{5.5}) to the (\ref{3.5}) is presented in detail in the appendix.

As in the case of the relativistic oscillator in absence of the external field~\cite{atakishiyev-wigner}, we can determine the equilibrium Wigner function for the relativistic oscillator in an external field by formula (\ref{3.8}) with the coefficients

\begin{equation}
\label{5.6}
w_n^g  = e^{ - \beta E_n^g } /Z^g (\beta ),
\end{equation}
where $E_n^g $ is the energy spectrum of the relativistic oscillator (\ref{4.4}) and the partition function has the following form:

\begin{equation}
\label{5.7}
Z^g (\beta ) = \sum\limits_{n = 0}^\infty  {e^{ - \beta E_n^g } }  = \frac{{e^{ - \beta \hbar \omega \delta \nu } }}{{1 - e^{ - \beta \hbar \omega \delta \nu } }}.
\end{equation}

Let us express the equilibrium Wigner function by the following series of the integrals:

\begin{eqnarray}
\label{5.8}
\fl W^g \left( {p,x} \right) = \frac{1}{{Z^g (\beta )}}\sum\limits_{n = 0}^\infty  {e^{ - \beta E_n^g } } W_n^g (p,x) \\ 
\fl  = \frac{{1 - e^{ - \beta \hbar \omega \delta } }}{{\pi \hbar }}\left( {2\delta \zeta } \right)^{2\nu } \sum\limits_{n = 0}^\infty  {\frac{{n!e^{\beta \hbar \omega \delta n} }}{{\Gamma \left( {n + 2\nu } \right)}}\int\limits_{ - \infty }^\infty  {e^{a\zeta e^y  + a^* \zeta e^{ - y} } L_n^{2\nu  - 1} \left( {2\delta \zeta e^y } \right)L_n^{2\nu  - 1} \left( {2\delta \zeta e^{ - y} } \right)e^{2ixy/\lambdabar } dy} }  .\nonumber
\end{eqnarray}

One can now use the following bilinear generating function for the generalized Laguerre polynomials~\cite{koekoek}

\begin{equation}
\label{5.9}
\fl \sum\limits_{n = 0}^\infty  {\frac{{n!z^n }}{{\Gamma \left( {n + \alpha  + 1} \right)}}L_n^\alpha  \left( x \right)L_n^\alpha  (y)}  = \frac{{(xyz)^{ - \alpha /2} }}{{1 - z}}I_\alpha  \left( {\frac{{2\sqrt {xyz} }}{{1 - z}}} \right)\exp \left( { - \frac{{z(x + y)}}{{1 - z}}} \right),\;\left| z \right| < 1
\end{equation}
and changing the integration and summation for the function $W^g \left( {p,x} \right)$ (\ref{5.8}) we obtain the following expression for the equilibrium Wigner function:

\begin{equation}
\label{5.10}
\fl W^g (p,x) = \frac{{4\delta \zeta }}{{\pi \hbar }}e^{-2\phi x/\lambdabar } e^{\beta \hbar \omega \delta \left( {\nu  - 1/2} \right)} I_{2\nu  - 1} \left( {\frac{{2\delta \zeta }}{{\sinh \frac{{\beta \hbar \omega \delta }}{2}}}} \right)K_{2ix/\lambdabar } \left( {2\zeta \sqrt {\rho ^2  + \delta ^2 \coth ^2 \frac{{\beta \hbar \omega \delta }}{2}} } \right),
\end{equation}
where $\phi  = \arg \left( {i\rho  + \delta \coth \frac{{\beta \hbar \omega \delta }}{2}} \right)$, $\rho  = \cos \varphi $ as well as $I_\alpha  \left( x \right)$ is the modified Bessel function of the first kind. To derive (\ref{5.10}), we have also used the formula (\ref{3.5a}).

It is obvious that the obtained expressions of the Wigner function for the relativistic oscillator in an external field in case of the $g=0$ (i.e., $\delta =1$ and $\rho = 0$) go to the corresponding expressions of the relativistic oscillator without external field~\cite{atakishiyev-wigner}.

%%%%%%%%%%%%%%%%%%%
\begin{figure}
\begin{center}
\begin{tabular}{cc}
\includegraphics[width=0.35\textwidth,angle=270]{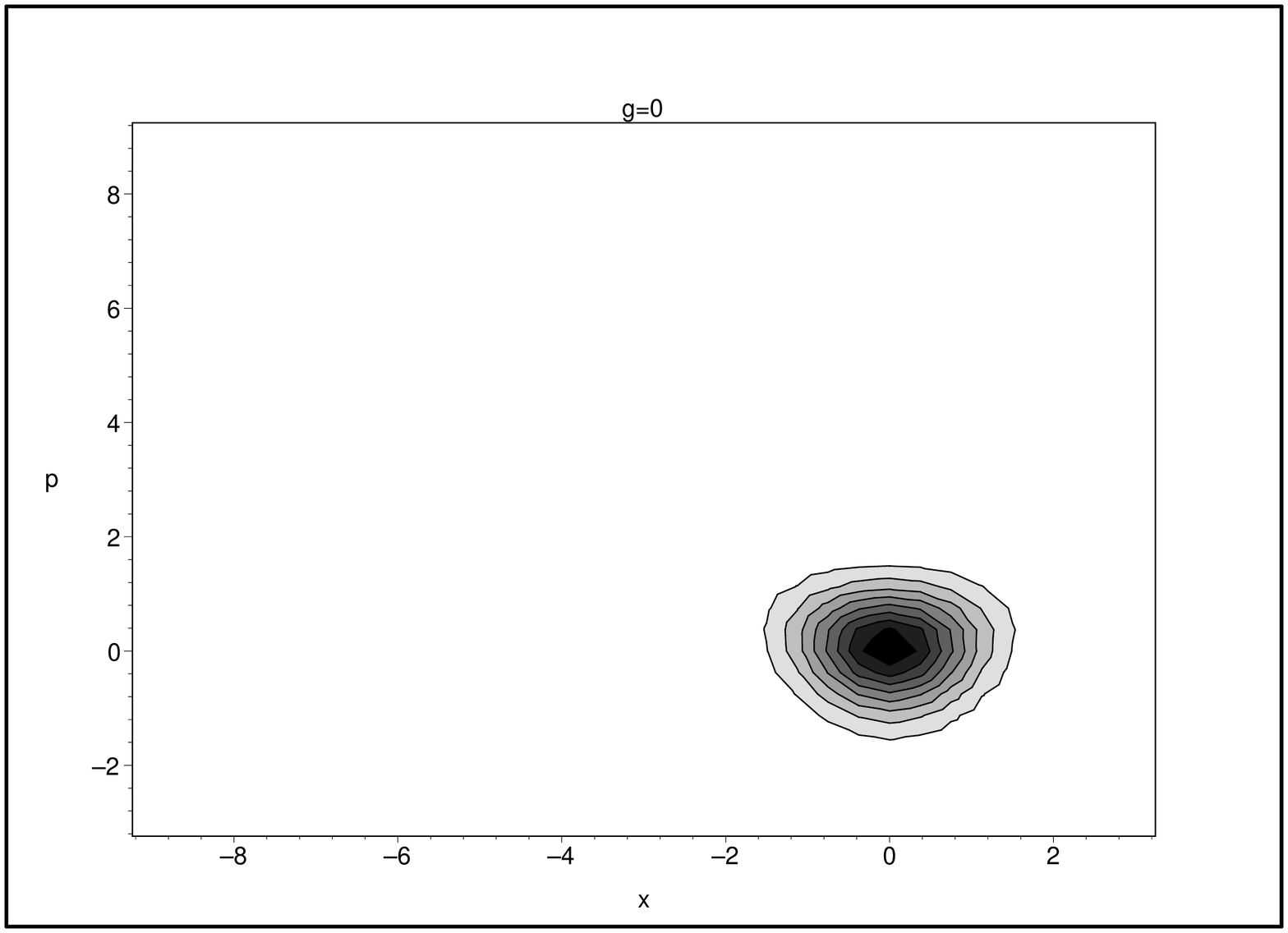}&
\includegraphics[width=0.35\textwidth,angle=270]{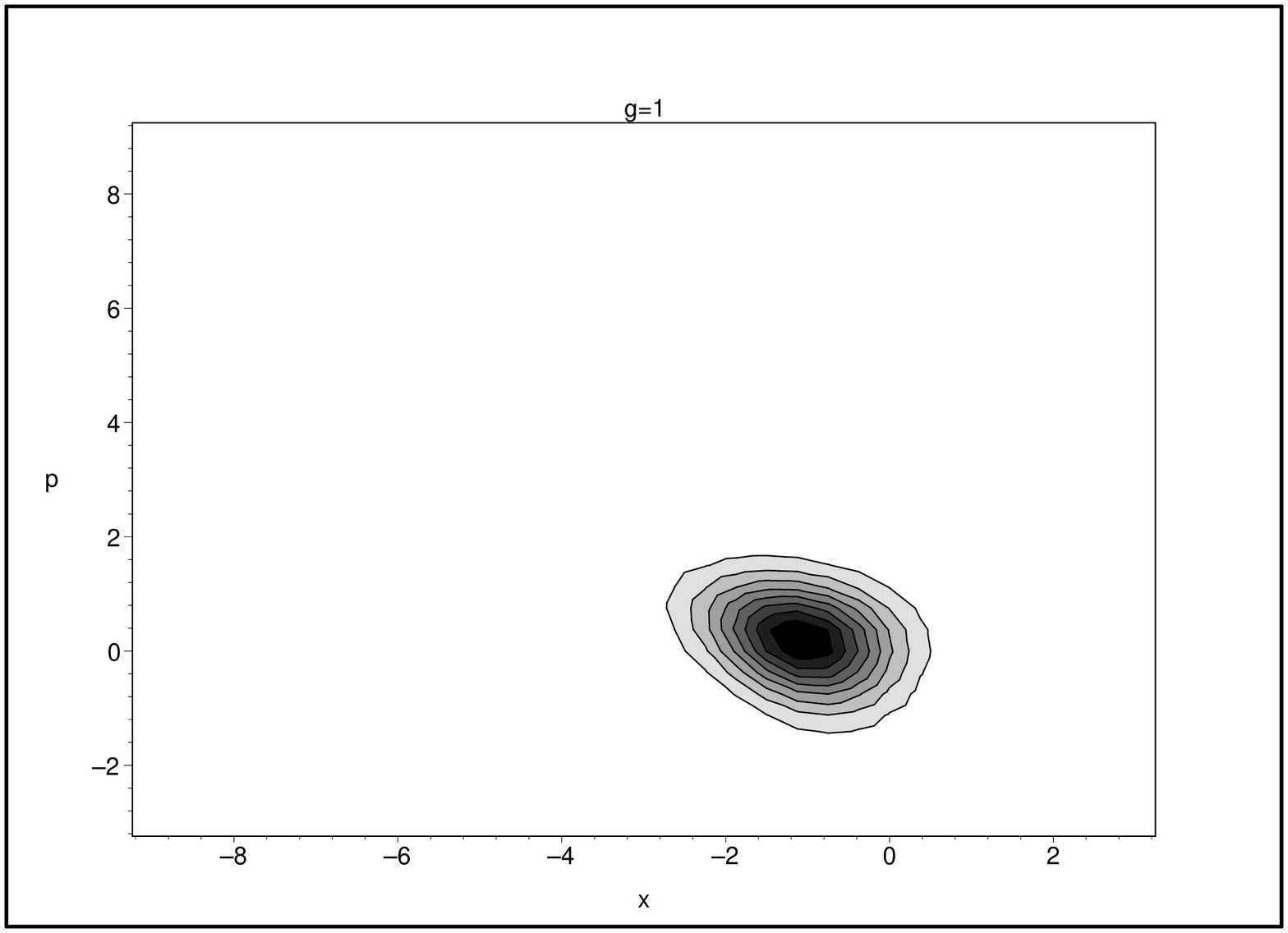}\\
\includegraphics[width=0.35\textwidth,angle=270]{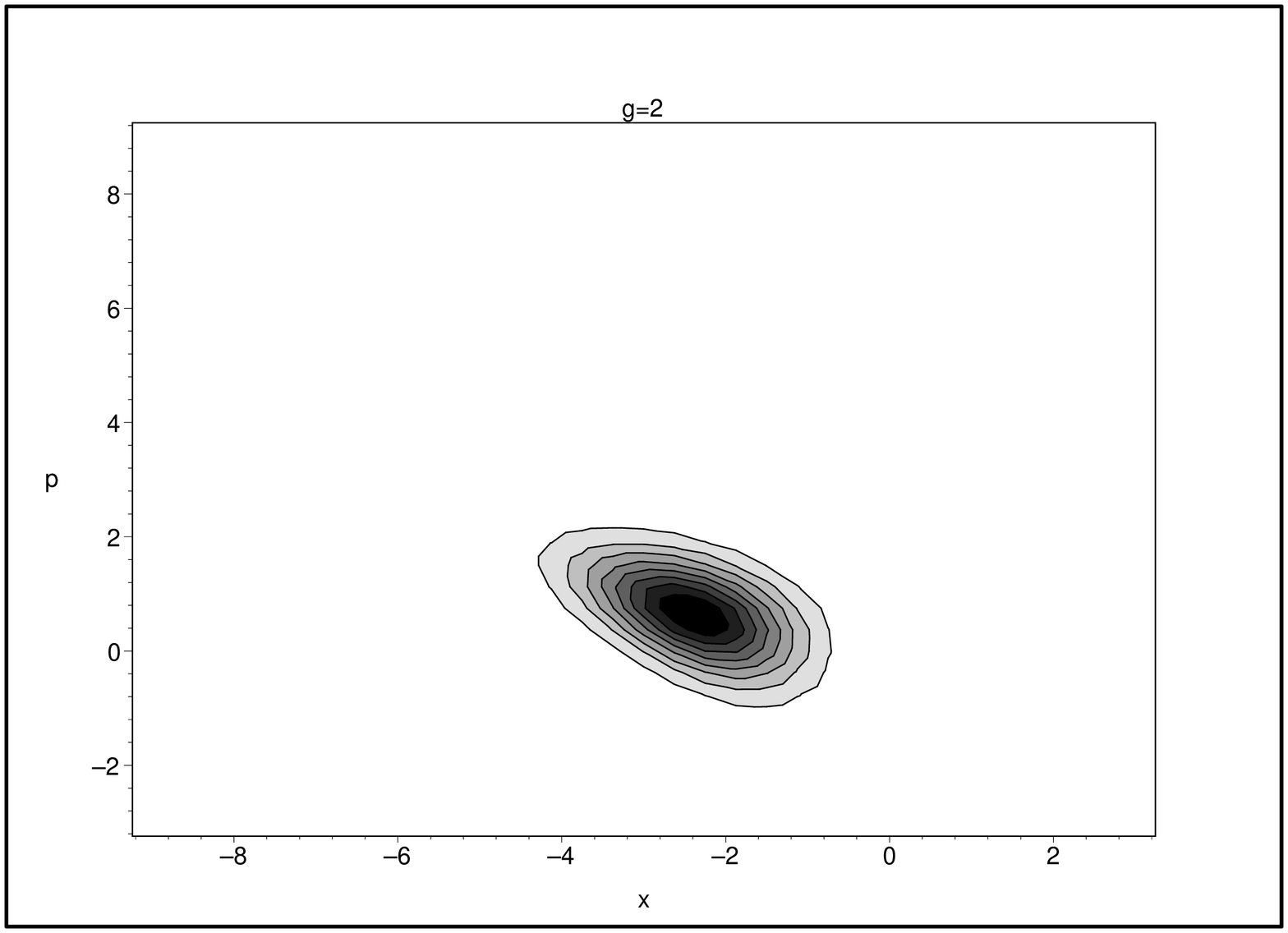}&
\includegraphics[width=0.35\textwidth,angle=270]{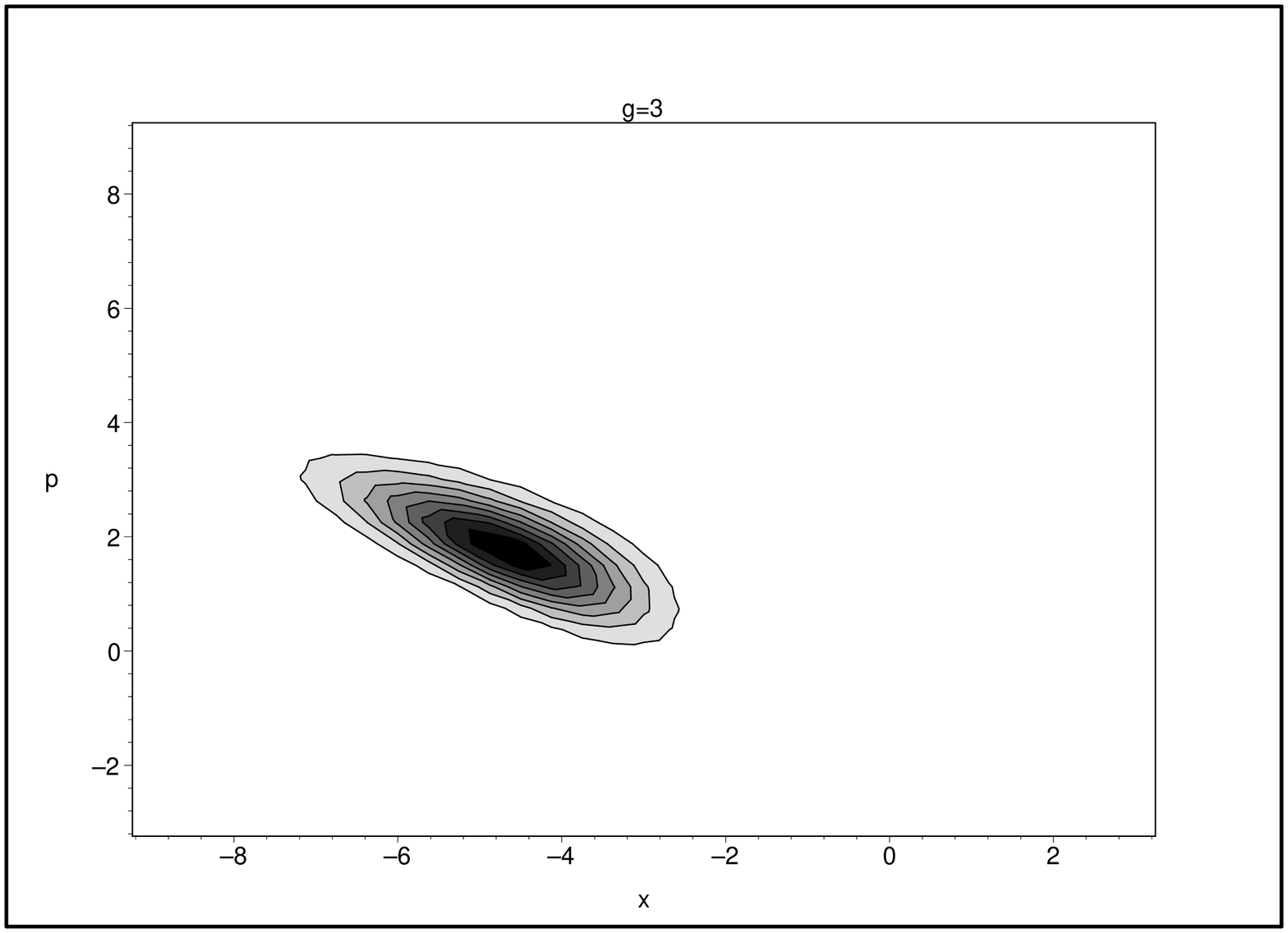}
\end{tabular}
\end{center}
\caption{\label{fig2} The behaviour of the ground-state Wigner function of the relativistic linear oscillator in an external homogeneous field for of speed of light $c= 4$ and values of the external field $g=0$, $1$, $2$ and $3$ ($m=\omega=\hbar=1$)}
\end{figure}
%%%%%%%%%%%

The behaviour of the ground-state Wigner function of the relativistic linear oscillator in an external homogeneous field $W_0^g (p,x)$ expressed through equation~(\ref{5.4}) is presented in \Fref{fig2}. One can mention here that the limit $c \rightarrow \infty$ recovers the non-relativistic case and behaviour of~(\ref{5.4}) will be identical to the non-relativistic ground state Wigner function $W_{N0}^g (p,x)$ determined by equation~(\ref{3.5}), which is presented in \Fref{fig1}. Then, any finite value of speed of light $c$ should show for us the contribution of the relativistic effects. Therefore, we depicted the joint distribution for the value of speed of light $c=4$ ($m=\omega=\hbar=1$) and see that unlike the non-relativistic case, more strong field to be compared with the speed of light cardinally changes its Gaussian-like form as well as shifts this distribution along the negative values of the position. One can see that the distribution in the absence of the field is determined for both positive and negative values of the momentum whereas a more strong field shifts it to the positive momentum values.

We do not present here graphical distribution of the Wigner function for the thermodynamic equilibrium state and the excited states of the relativistic oscillator under influence of the external field. Only we need to note here that the behaviour of the relativistic oscillator excited states in the phase space is similar to the ground-state Wigner function behaviour. Same picture can be observed for the Wigner function for the thermodynamic equilibrium states, which is a good approximation of the ground state joint distribution for the room temperature.

\appendix
\section{}
\vskip 0.5cm

In this appendix, we compute the non-relativistic limit of the Wigner functions of the relativistic linear oscillator in an external field. First, we have to take into account that

\[
\fl\qquad \rho  = \cos \varphi  = \frac{{\xi _0 }}{{\sqrt \mu  }},\;\delta  = \sin \varphi  = \sqrt {1 - \frac{{\xi _0 ^2 }}{\mu }} ,\;x/\lambdabar=\sqrt \mu \xi,\;\zeta  = \sqrt \mu  \left( {\eta  + \sqrt {\mu ^2  + \eta ^2 } } \right),
\]
\[
\nu  = \frac{1}{2} + \sqrt {\frac{1}{4} + \mu ^2 } ,\;\mu  = \frac{{mc^2 }}{{\hbar \omega }}.
\]

Then, at $\mu  \to \infty $ we will have the following limit relations for the elementary functions:

\begin{eqnarray}
\label{a.1}
 \mathop {\lim }\limits_{\mu  \to \infty } 2\phi x/\lambdabar  = \frac{{2\xi \xi _0 }}{b}, \nonumber \\
 \mathop {\lim }\limits_{\mu  \to \infty } \left( {2\varphi  - \pi } \right)x/\lambdabar  =  - 2\xi \xi _0 , \\ 
 \left( {2\delta \zeta } \right)^{2\nu }  \cong e^{2\mu \ln 2\mu  + 2\sqrt \mu  \eta  - \xi _0 ^2 } , \nonumber
\end{eqnarray}
as well as for the special functions:

\begin{eqnarray}
\label{a.2}
 \mathop {\lim }\limits_{\mu  \to \infty } \nu ^{ - \frac{n}{2}} L_n^{2\nu  - 1} \left( {2\delta \zeta e^{\frac{{i\lambdabar }}{2}\partial _x } } \right) = \frac{{\left( { - 1} \right)^n }}{{n!}}H_n \left( {\eta  + \frac{i}{2}\partial _\xi  } \right), \nonumber \\
I_{2\nu  - 1} \left( {2\delta a\zeta } \right) \cong \frac{1}{{\sqrt {4\pi \mu b} }}e^{2\mu \left( {b - f } \right) + 2\sqrt \mu  b \eta  + \left( {b' - \frac{1}{{b}}} \right)\eta ^2  - b\xi _0 ^2 } , \\ 
K_{2ix/\lambdabar } \left( {2\tilde b \zeta } \right) \cong \sqrt {\frac{\pi }{{4\mu \tilde b }}} e^{ - 2\mu \tilde b  - 2\sqrt \mu  \tilde b \eta  - \tilde b \eta ^2  - \frac{{\xi ^2 }}{{\tilde b }}} , \nonumber
\end{eqnarray}
where:

\[
a = \frac{1}{{\sinh f}},\;b = \coth f ,\;f  = \frac{{\beta \hbar \omega \delta }}{2},\;\tilde b  = \sqrt {\rho ^2  + \delta ^2 b^2 } ,\;\tilde b  \cong b + \frac{{\xi _0 ^2 }}{{2\mu }}\left( {\frac{1}{{b}} - b} \right).
\]

To obtain the formulae (\ref{a.2}) we used the following representations for the modified Bessel, the Macdonald and the gamma functions~\cite{bateman}:

\begin{eqnarray}
 \sqrt {2\pi } I_p \left( x \right) \cong \frac{{e^{\sqrt {p^2  + x^2 } - p\sinh ^{ - 1} \frac{p}{x} } }}{{\sqrt[4]{{p^2  + x^2 }}}},\quad x \approx p \to \infty , \nonumber \\ 
 K_{ip} \left( x \right) \cong \sqrt {\frac{\pi }{2}}  \cdot \frac{1}{{\sqrt[4]{{x^2  - p^2 }}}} \cdot e^{ - \sqrt {x^2  - p^2 }  - p \cdot \sin ^{ - 1} \frac{p}{x}} ,\;x > p > 0,\;p \to \infty , \nonumber \\ 
 \Gamma \left( z \right) \cong \sqrt {\frac{{2\pi }}{z}}  \cdot e^{z\ln z - z} ,\;\left| z \right| \to \infty , \nonumber 
\end{eqnarray}
as well as one needed to take into account the following limit relation between the Laguerre and the Hermite polynomials~\cite{szego}:

\[
\mathop {\lim }\limits_{\nu  \to \infty } \nu ^{ - \frac{n}{2}} L_n^{2\nu  - 1} \left( {2\nu  + 2\sqrt \nu  x} \right) = \frac{{\left( { - 1} \right)^n }}{{n!}}H_n \left( x \right).
\]

By using the equations (\ref{a.1}) and (\ref{a.2}), one can show that at $c \to \infty$, expressions (\ref{5.2}) and (\ref{5.10}) coincide with the expressions (\ref{3.5a}) and (\ref{3.10}), respectively.


\section*{References}

\begin{thebibliography}{99}

\bibitem{moshinsky}
Moshinsky M 1969 {\it The Harmonic Oscillator in Modern Physics: from Atoms to Quarks} (New-York: Gordon and Breach)

\bibitem{landau}
Landau L D and  Lifshitz E M 1997 {\it Quantum Mechanics: Non-Relativistic Theory} (Oxford: Butterworth-Heinemann)

\bibitem{agarwal}
Agarwal G S 1971 \PR {\bf A4} 739

\bibitem{wallis}
Wallis H, Dalibard J and Cohen-Tannoudji C 1992 {\it App. Phys.} {\bf B54} 407

\bibitem{kulikov}
Kulikov I K 2002 {\it Int. J Theor. Phys.} {\bf 41} 1281

\bibitem{saif}
Saif F 2005 {\it Phys. Rep.} {\bf 419} 207

\bibitem{azevedo1}
Azevedo S 2001 {\PL} {\bf A288} 33

\bibitem{azevedo2}
Azevedo S 2005 {\it Int. J Quantum Chem.} {\bf 101} 127

\bibitem{atakishiyev-nonlin}
Mir-Kasimov RM, Nagiyev S M and Kagramanov E D 1987 {\it The relativistic linear oscillator in a homogeneous external field and the bilinear generating functions for the Pollaczek polynomials, Preprint No 214} (Institute of Physics, Azerbaijan Academy of Sciences) 12p

\bibitem{atakishiyev-adv}
Atakishiyev N M, Nagiyev S M and Wolf K B 1996 {\it Advanced Series in Nonlinear Dynamics, Vol. 8: New Trends for Hamiltonian Systems and Celestial Mechanics} p. 15 (Singapore: World Scientific)

\bibitem{atakishiyev-rep}
Atakishiyev N M and Wolf K B 1989 {\it Rep. Math. Phys.} {\bf 27} 305

\bibitem{wigner}
Wigner E P 1932 {\PR} {\bf 40} 749

\bibitem{lee}
Lee H W 1995 {\it Phys. Rep.} {\bf 259} 147

\bibitem{hillery}
Hillery M, O'Connell R F, Scully M O and Wigner E P 1984 {\it Phys. Rep.} {\bf 106} 121

\bibitem{tatarskii}
Tatarskii V I 1983 {\it Sov. Phys. Uspekhi} {\bf 26} 311

\bibitem{davies}
Davies R W and Davies K T R 1975 \APNY {\bf 89} 261

\bibitem{akhundova}
Akhundova E A, Dodonov V V and Manko V I 1982 {\it Physica} {\bf A115} 215

\bibitem{atakishiyev1}
Atakishiyev N M, Mir-Kasimov R M and Nagiyev S M 1980 {\it Theor. Math. Phys.} {\bf 44} 47

\bibitem{atakishiyev-wigner}
Atakishiyev N M, Nagiyev S M and Wolf K B 1998 {\it Theor. Math. Phys.} {\bf 114} 322

\bibitem{prudnikov-I}
Prudnikov A P, Brychkov Y A and Marichev O I 1986 {\it Integrals and Series - vol 1: Elementary Functions} (New-York: Gordon\&Breach)

\bibitem{shapiro}
Shapiro I S 1956 {\it Sov. Phys.-Dokl.} {\bf 1} 91

\bibitem{alonso}
Alonso M A, Pogosyan G S and Wolf K B 2002 \JMP {\bf 43} 5857

\bibitem{koekoek}
Koekoek R and Swarttouw R F 1998 {\it The Askey-scheme of hypergeometric orthogonal polynomials and its q-analogue} (Delft University of Technology: Report no. 98-17)

\bibitem{bateman}
Bateman H and Erdelyi A 1955 {\it Higher Transcendental Functions - vol 2} (New-York: McGraw-Hill)

\bibitem{szego}
Szeg\"o G 1975 {\it Orthogonal Polynomials, ed. R.I. Fourth} (Providence: American Math. Soc.)

\end{thebibliography}
\end{document}